# Phase Diagram for Splay Glass Superconductivity

T. P. Devereaux, R. T. Scalettar, G. T. Zimanyi, K. Moon

*Department of Physics, University of California, Davis, CA 95616*

E. Loh

*Thinking Machines Corp., 245 First Street, Cambridge, MA 02142*

## Abstract

Localization of flux lines to splayed columnar pins is studied. A sine-Gordon type renormalization group study reveals the existence of a Splay glass phase and yields an analytic form for the transition temperature into the glass phase. As an independent test, the $I-V$ characteristics are determined via a Molecular Dynamics code. The glass transition temperature supports the RG results convincingly. The full phase diagram of the model is constructed.

Typeset using REVTEX

Zero resistance in Type $II$ superconductors is maintained in high magnetic fields by pinning the flux lines to avoid dissipation. The central issue of technological interest is how to maximize the critical current $J_c$ which depins the vortices. Atomic vacancies or light particle irradiation typically modelled by random point disorder collectively pin the flux lines into an entangled arrangement leading to the formation of the Vortex glass phase [1]. On the other hand, columnar pins produced for example by heavy ion irradiation, twinning, or screw dislocations give rise to the Bose glass phase [2,3]. It was argued that $J_c$ is strongly enhanced by columnar pinning over the traditional point defects [2]. The two types of glasses are characterized by distinct localization mechanisms, manifested in different $I$-$V$ curves. Recently the phase diagram of the model with the simultaneous presence of both types of disorder has been constructed [4]. The main result was that the nature of the localization is dominantly of the Bose glass type with a reduced transition temperature [5].

In the quest of maximizing $J_c$, the idea of tilting the columnar defects was proposed [6]. In the case of parallel columns, once a segment of the vortex reaches an adjacent column either via thermal activation or a quantum mechanical process, the remaining part of the vortex can follow at no additional energy cost. In the Splay glass phase, during a vortex hop an ever increasing segment of the line will be forced into an energetically unfavorable region. This provides an effective confining mechanism and leads to a dramatically enhanced value of $J_c$. Subsequent experiments have indeed confirmed this expectation [7]. In spite of intense interest, the phase diagram of the model has been determined only qualitatively. The main result of the present work is the quantitative determination of the phase diagram. First a renormalization group (RG) analysis is used to determine the dependence of $T_c$ on the degree of columnar tilting and flux line rigidity. This RG treatment is then shown to be in excellent agreement with molecular dynamics simulations. From a theoretical point of view the problem is also interesting because of its relevance to Quantum Phase transitions occuring in boson systems [3,8].

Our analysis begins with the Hamiltonian for $N$ interacting flux lines confined to a plane,

$$H = \int dz \sum_{i=1}^{N} \left\{ \frac{1}{2} \tilde{\epsilon}_1 \left[ \frac{dr_i(z)}{dz} \right]^2 + \frac{\nu_0}{2} \sum_{j \neq i} \delta[r_i(z) - r_j(z)] + V[r_i(z), z] \right\}. \tag{1}$$

Here the scalar displacement field $r_i(z)$ describes the position of the *i-th* flux line in a direction perpendicular to the applied field $(z)$, $\nu_0$ denotes the strength of the line-line interaction, $V$ denotes the disorder potential, and $\tilde{\epsilon}_1$ the elastic energy cost per unit length for transverse wandering. We choose to work with this restricted model since it is more amenable to analytic and numerical treatment than the corresponding 3-D model. It has been recently suggested via scaling arguments that the fluctuations of the flux line along the direction of the applied current are irrelevant at the transition and thus a planar model may be immediately applicable to 3-D systems [9].

The correlator $\Delta$ of the disorder potential $V$ is defined via

$$\langle V(x,z)V(x',z') \rangle = \Delta(x - x', z - z'). \tag{2}$$

For the case of uncorrelated point disorder, $\Delta_p(x,z) \sim \delta(x)\delta(z)$. Columnar disorder is correlated along the ion trajectories. When aligned parallel to the applied magnetic field $\Delta_c(x,z) \sim \delta(x)$. These two cases have been analyzed separately in Refs. [1–3], while the competition between the two types of disorder was investigated in Refs. [4,5]. In the present work, the disorder correlator represents tilted columnar lines with a Gaussian distribution of tilting angles of width $v_D$. The Fourier transform of the correlator is

$$\Delta_s(q_x, q_z) = \Delta_0 \frac{\Lambda_x}{\sqrt{2\pi} v_D |q_x|} exp[-(q_z \Lambda_x)^2 / 2 v_D^2 (q_x \Lambda_z)^2]$$
$$= \Delta_0 h_s(q_x, q_z), \tag{3}$$

as was also used in Ref. [6]. Here $\Lambda_{x,z}$ are the ultraviolet cut-offs and we have neglected the width of the vortex core. For the case of $v_D = 0$, $\Delta$ recovers the untilted columnar form, while $v_D \to \infty$ describes an isotropic distribution of tilting angles.

The Hamiltonian can be rewritten in a sine-Gordon-like form as

$$H/T = \sum_{\alpha,\beta} \left\{ \int dq_x dq_z \frac{1}{2} \left[ \sum_{i=x,z} K_i q_i^2 \delta_{\alpha,\beta} - q_z^2 \left( \frac{K_z}{K_x} \right)^2 \Delta'(q_x, q_z) \right] u_\alpha(q_x, q_z) u_\beta(q_x, q_z) \right. \tag{4}$$
$$\left. - g \int dx dx' dz dz' h_s(x', z') \cos[2\pi(u_\alpha(x,z) - u_\beta(x + x', z + z'))] \right\}.$$

Here $u$ is the displacement field in a continuum description. We have applied a coarse-graining procedure and introduced replicas (indexed by $\alpha, \beta = 1, \cdots, n$) to handle the disorder. The parameters $K_x = \nu_0/T$ and $K_z = a\tilde{c}_1/T$ are the bulk and tilt moduli, respectively, which describe the elasticity of the flux line array, and $\Delta_x = \Delta_0/T^2$. The bare coupling constant of the nonlinear term, representing the interactions, is given by $g = \Delta_0/(aT)^2$, with $a$ the average line separation. As shown in Refs. [4,5], the transformation $u(x,z) \to u(x,z) + \frac{1}{\nu_0}\int^x d\bar{x} V(\bar{x}, z)$ before replicating Eq. (4) allows us to absorb the forward scattering in the $x$ direction $\propto \partial_x u$ at the expense of generating a forward scattering in the $z$ direction $\propto \partial_z u$. We thus obtain

$$\Delta'(q_x, q_z) = \Delta_x \frac{\Lambda_x^3}{\Lambda_z^2 \sqrt{2\pi} v_D} \frac{q_z^2}{|q_x|^3} \, exp[-(q_z \Lambda_x)^2 / 2v_D^2 (q_x \Lambda_z)^2]. \tag{5}$$

This has the appealing feature that in the case of no tilting ($v_D = 0$), the Bose glass transition is immediately recovered [2,3].

We have performed a renormalization-group analysis of the model Eq. (4) to first order in the interaction $g$. We find the following recursion relations upon rescaling by a factor $e^l$ and taking the $n \to 0$ limit:

$$\frac{dK_x}{dl} = 0 + O(g^2), \tag{6}$$

$$\frac{dK_z}{dl} = \begin{cases} 0, & \text{for } v_D \neq 0, \\ C_1 g, & \text{for } v_D = 0, \end{cases} + O(g^2), \tag{7}$$

$$\frac{d\Delta_x}{dl} = C_2 g^2, \tag{8}$$

$$\frac{dg}{dl} = g\left[3 - K^{-1}\left[1 + \frac{3\Delta_x}{4\pi K_x}\sqrt{\alpha} f(1/v_D^2 \alpha)\right]\right] + O(g^2). \tag{9}$$

Here $K^{-1} = 2\pi(K_x K_z)^{-1/2} \sim T$ and the ratio of the tilt to bulk modulus, $\alpha = K_z/K_x = \tilde{c}_1 a/\nu_0$ describes the elastic anisotropy of the flux array, $C_{1,2}$ are positive constants, and $f$ is a function to be specified below. The bulk modulus $K_x$ is unrenormalized at first order (unlike the case of point or columnar disorder, our model does not possess statistical invariance to tilt [4]), while the renormalization of the tilt modulus $K_z$ only occurs for no tilting. This distinguishes the Bose glass ($v_D = 0$) from the Splay glass, ($v_D \neq 0$). In the former, the tilt modulus scales to infinity, while in the latter it remains finite both above and below the transition. This suggests that the Bose glass is indeed unstable to the splaying of the columns and is replaced by a new thermodynamic phase. In fact, in many respects the recursion relations are similar to those obtained for the case of point disorder. Hence it is also termed "correlated Vortex glass" [6].

The most important feature of the recursion relations is the reduction of the glass transition temperature not only for increased tilt of the columns but also for increased elastic anisotropy of the flux line array. The Splay glass transition temperature is given by

$$K_{S.G.}^{-1} = K_{B.G.}^{-1} \left[ 1 + \frac{3\Delta_x}{4\pi K_x} \sqrt{\alpha} f(1/v_D^2 \alpha) \right]^{-1}. \tag{10}$$

The function $f$ can be written in terms of a confluent hypergeometric function $U$ [10],

$$f(x) = U(2, 1/2, x/2).$$

It has the limiting behavior $f(x \to 0) = 4/3 + O(\sqrt{x})$, and $f(x \to \infty) = 4/x^2 - 40/x^3 + O(1/x^4)$. For the case $v_D = 0$, the previous results [2,3] for the Bose glass are recovered, $K_{B.G.}^{-1} = 3$. For finite tilting the localization transition occurs at a lower temperature, in agreement with the arguments made in Ref. [6]. The phase boundary is shown in Fig. (1) as a function of the width $v_D$ of the tilting angle distribution and the anisotropy parameter $\alpha$. For small tilting we find,

$$K_{S.G.}^{-1}/K_{B.G.}^{-1} = 1 - \frac{1}{\pi} \frac{\Delta_x}{K_x} \alpha^{5/2} v_D^4 + O(v_D^6). \tag{11}$$

The transition temperature falls off more slowly with tilting angle ($\sim v_D^4$) than suggested from the arguments presented in Ref. [6]. From an application point of view, it is worth noting that the reduction of the transition temperature can be counteracted by increasing the elasticity anisotropy, as represented in Eq. (9) by the parameter $\alpha$. This feature is absent for the case of straight columnar pins. Lastly we note that for fixed $\alpha$, the glass transition temperature saturates at a non-zero value for the limit of $v_D \to \infty$, i.e. the isotropic limit. In this limit the columnar pins become more and more point-like and thus are relatively ineffective at pinning, leading to a correspondingly lower transition temperature.

To check the predictions of the RG, we have performed molecular dynamics simulations of Eq. (1) directly. The most direct way to identify the glass phase is based on the study of the response of the flux lines to an external current [11,12]. Assuming overdamped dynamics, the equation of motion for the lines is given by

$$\partial_t r = \eta \partial_x^2 r - \partial_x (V_p + V_v) + F + f. \qquad (12)$$

Here $\eta$ sets the energy scale for bending of a line ($\eta \propto \tilde{\epsilon}_1$), $V_{p,v}$ are the pinning and vortex interaction potentials, $F$ is the Lorentz force exerted by the applied current, and $f$ is the Langevin noise taken to model the effects of temperature $T$ via

$$\langle f(x,z,t) f(x',z',t') \rangle = 2T \delta(x-x') \delta(z-z') \delta(t-t'). \qquad (13)$$

The potentials are chosen as follows: (1) The disorder pinning potential is modeled as the smooth function $V_p(x,z) = -V_{pin}[(x/R_{pin} - 1)(x/R_{pin} + 1)]^2$ for $|x| < R_{pin}$ and 0 otherwise, where $x$ measures the distance from the center of the pin, given by $x_0 = vz$, to the line and $v$ is the tangent of the angle of the splayed columns which is chosen randomly from a uniform distribution of width $v_D$. The position of the columns are randomly placed in the $x-z$ plane. (2) The vortex interaction potential is taken to be $V_v = V_{vortex} e^{-x/R_{vortex}}$ for $|x| < 6 R_{vortex}$ and 0 otherwise, where x is the vortex-vortex separation. All length scales are measured in terms of $R_{vortex}$ which is given approximately by the screening length $\lambda \sim 5000$ angstroms, while energy scales are set by $V_{vortex} = 2(\Phi_0/4\pi\lambda)^2$, with $\Phi_0$ the flux quantum. We lastly chose to work with narrow pinning defects so that $R_{pin} = 0.3 R_{vortex}$.

The parameters for our simulations were chosen as follows: our system size was $32 \times 32$ in units of $R_{vortex}$, and contained 20 lines and 32 columns. The parameters $\eta = 2.5, V_{pin} = V_{vortex} = 1$ were fixed for all runs. While non-unique, this choice of the parameters falls within the range of values pertinent to cuprate superconductors [13,8]. The average velocity of the flux line array was measured for different temperatures and averaged over a varying number of disorder configurations so as to keep the statistical fluctuations below half a percent. Using $1/V_{vortex}$ to set the time scale, typically less than 2,000 cycles were sufficient to reduce the velocity fluctuations to values less than 1 percent, thus ensuring proper equilibration of the system. A typical series of results are shown in Fig. (2) in a log-log plot. Here the tilting angle distribution width was $V_D = 2$. Two regions can be clearly seen at low temperatures - a linear response at large driving forces corresponding to free flux motion, and a non-linear response corresponding to thermally assisted flux motion. The crossover between the two becomes broader as the temperature is increased and the Ohmic region becomes larger.

We define the Splay glass transition temperature as the temperature at which the entire curve obeyed an Ohmic response [14]. Our results for different tilting angles are shown in Fig. (3). Other methods for establishing the critical temperature, such as identifying the point of maximum slope of the $I - V$ curves or monitoring when the fluctuations of the velocity due to the disorder drop below 1 percent, yield the same results to within the error bars. The solid line displays the transition temperature obtained from the RG method. The nonuniversal Bose glass transition temperature, which is 3 in the RG treatment, depends on the microscopic details of the model. In our numerical study $T_{B.G.}$ was determined from simulations with no tilting of the columnar defects. With the above choice of parameters we obtained $K_{B.G.}^{-1} = 2.35 \pm 0.05$. The solid line in Fig. (3) is subsequently obtained by inserting this value into Eq. (10). No adjustable parameters enter the expression. We have explored a larger region of the parameter space by changing the time step $dt$, $\eta$, $V_{vortex}/V_{pin}$, and the line density, and found that these changes modified the overall transition temperature only through the combination defined by $\alpha$ as suggested by the functional form obtained from the RG treatment. The numerical data support the RG results remarkably well lending credence to the full phase diagram of the model given in Fig. 1. This is the central result of our paper.

In conclusion, we have investigated the localization transition of flux lines in the presence of tilted columnar defects using both RG and molecular dynamics methods. We find that the Splay glass transition temperature decreases for increased tilting of the columnar pins. The functional form for the transition temperature derived from an RG treatment is convincingly confirmed by our numerical simulations. We note that the experiments of Ref. [7] were only performed for small angle splay, $v_D \sim 0.176$, and thus it would be a useful check of the theory to repeat these experiments with larger degree of splay. A full finite size scaling analysis is needed to determine the critical properties of the Splay glass transition.

We wish to thank G. Batrouni, T. Hwa, T. Imbuk III, and A. Zawadowski for discussions. This work was supported by N.S.F. Grant No. 92-06023. GTZ was also funded by the donors of the Petroleum Research Fund administered by the American Chemical Society. K. Moon was supported by a LACOR grant from Los Alamos National Laboratory.

[13] See the review articles in *Phenomenology and Applications of High Temperature Superconductors*, edited by K. Bedell *et al.*, Addison- Wesley, Reading, MA (1992).

[14] The true critical temperature must be determined via a finite size scaling analysis as performed in Refs. [12]. However we are not aware of any predictions for the scaling behavior. Since our primary goal is only to extract the tilting angle dependence of the critical temperature, this approximate definition should be sufficient.

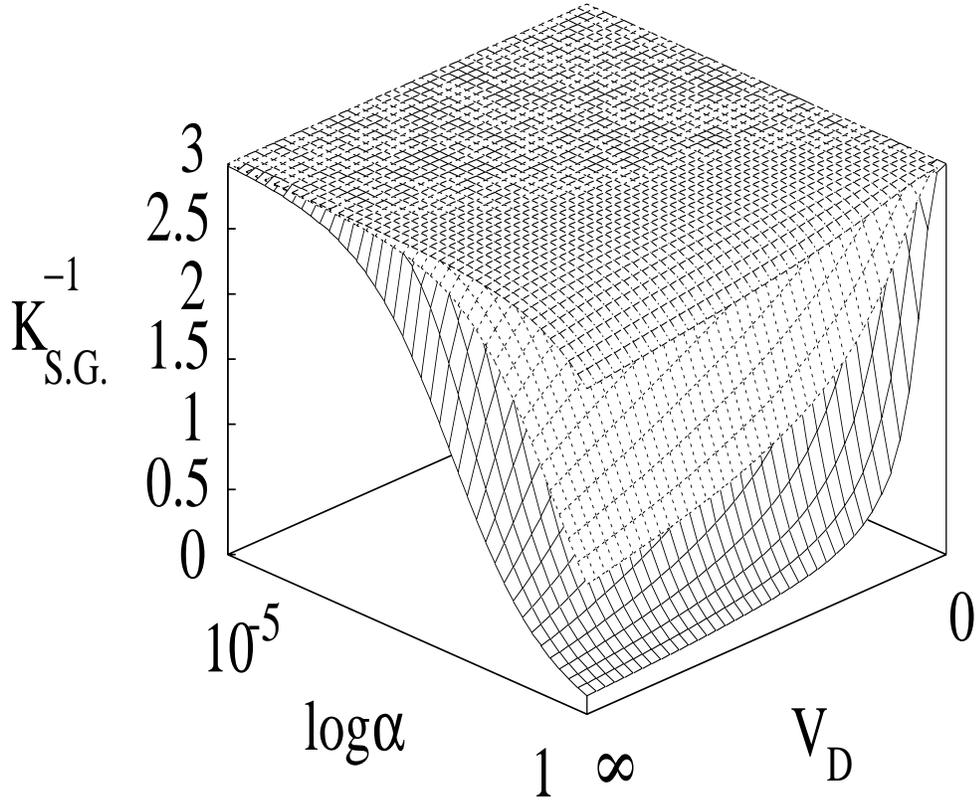

FIG. 1. Splay glass transition temperature as a function of the width $v_D$ of the tilting angle distribution and the elastic anisotropy parameter $\alpha = \tilde{\epsilon}_1 a/\nu_0$, for various values of $\Delta_x/K_x$. The top, middle, bottom surfaces are for $\Delta_x/K_x = 0.2, 2$, and $20$, respectively.

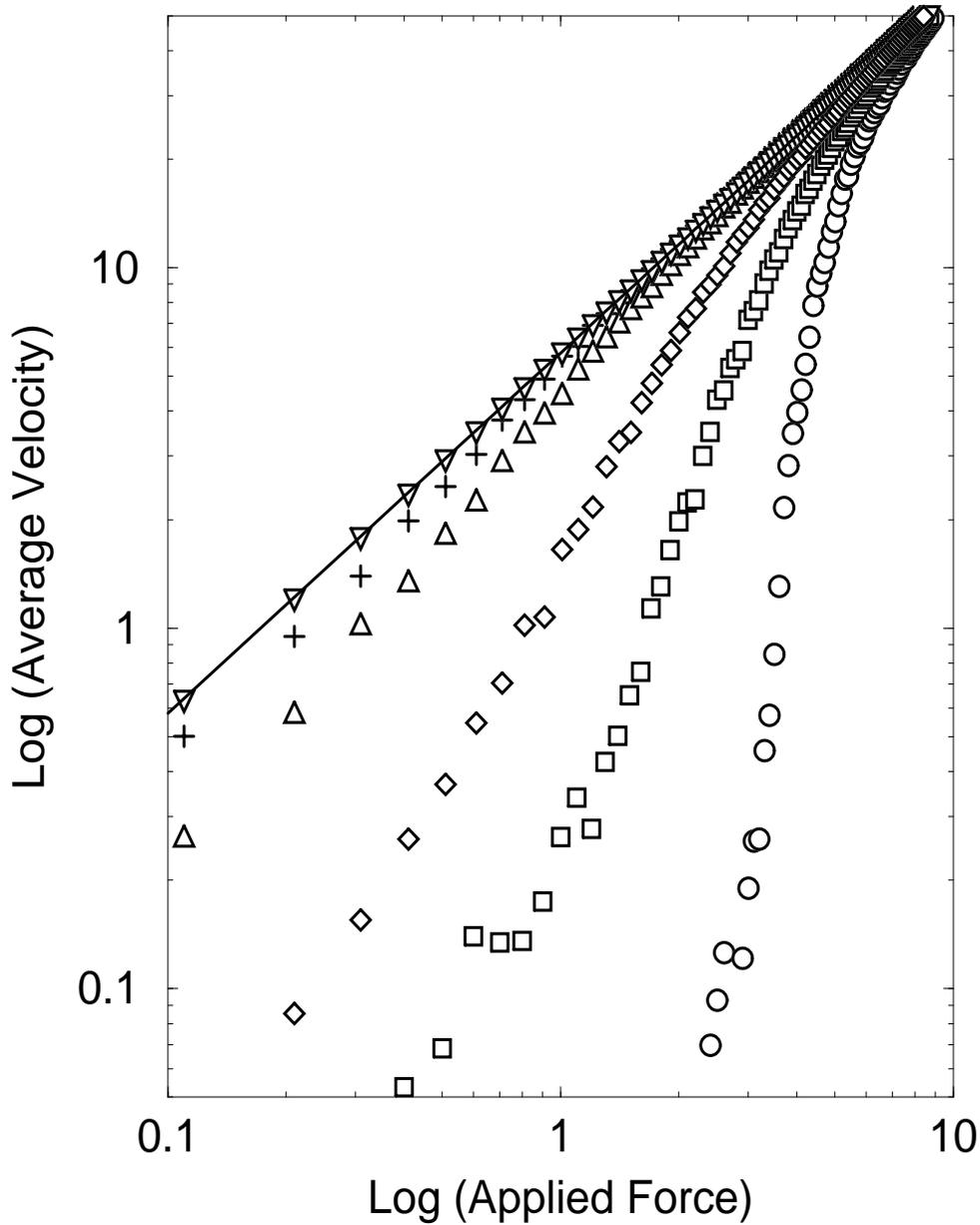

Devereaux et al, Fig. 2

FIG. 2. Average velocity of flux line motion versus applied force for T=0.01 (circles), 0.1 (squares), 0.26 (diamonds), 1.26 (up triangles), 1.81 (pluses), and 1.91 (down triangles). The solid line represents Ohmic behavior (slope of 1) and serves as a guide to the eye. The critical temperature is estimated to be 1.87. Runs were performed for $V_D = 2$ and averaged over 10 disorder configurations. Here the statistical fluctuations are smaller than the size of the points and $\alpha = 4$ is kept fixed.

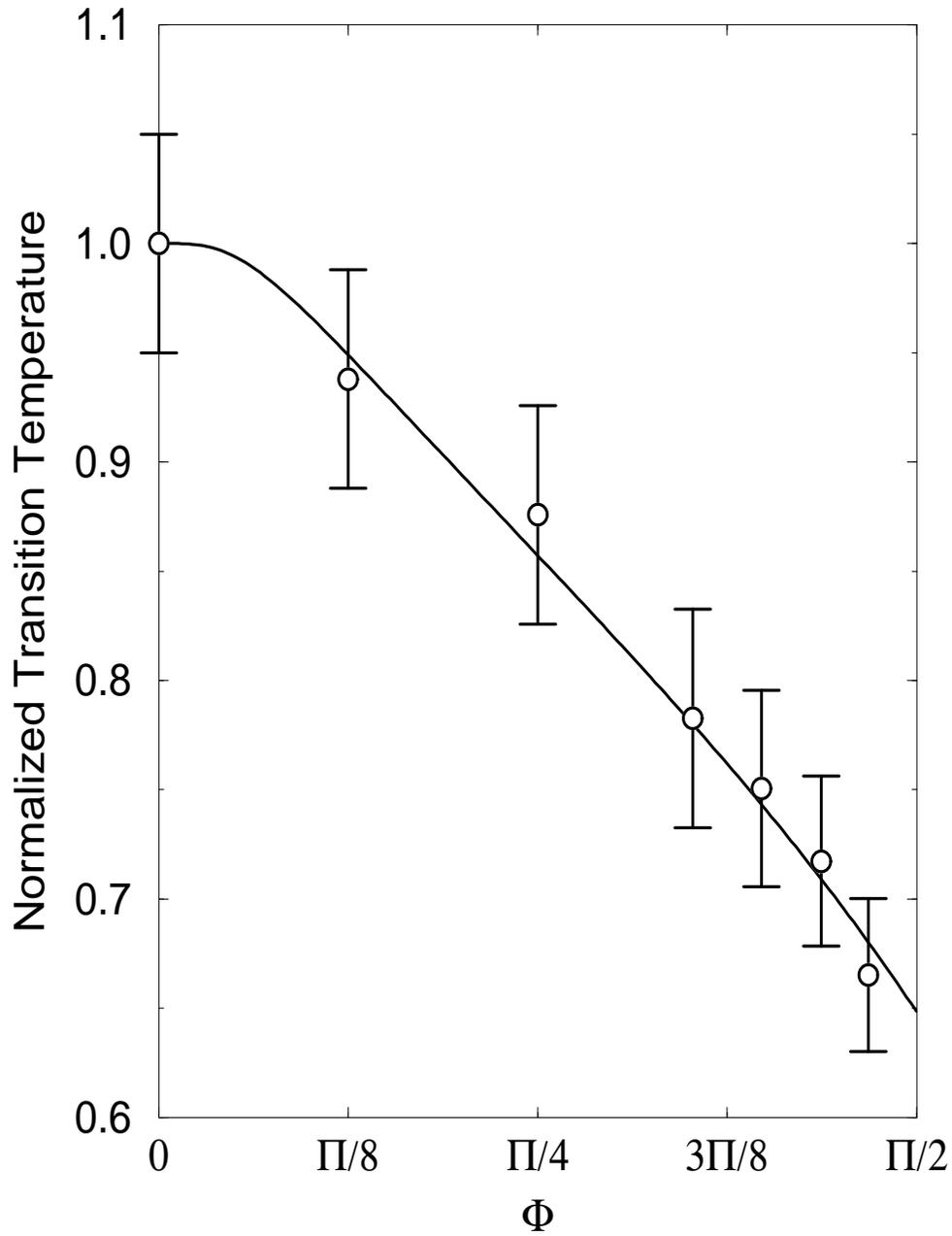

Devereaux et al, Fig. 3

FIG. 3. Results for the Splay glass phase transition temperature derived from molecular dynamics simulations of Eq. (11). The circles represent runs for the tilting angle distribution widths $v_D = \tan(\Phi)$, while keeping $\alpha = 4$ fixed.